\documentclass[preprint2]{emulateapj}

\usepackage{gensymb}

\shorttitle{DCO$^+$ in IM Lup}
\shortauthors{\"Oberg et al.}

\begin{document}

\title{Double DCO$^+$ rings reveal CO ice desorption in the outer disk around IM Lup}

\author{Karin I. \"Oberg}
\affil{Harvard-Smithsonian Center for Astrophysics, 60 Garden St., Cambridge, MA 02138}

\author{Kenji Furuya}
\affil{Leiden Observatory, Leiden University, PO Box 9513, 2300 RA, Leiden, The Netherlands}

\author{Ryan Loomis}
\affil{Harvard-Smithsonian Center for Astrophysics, 60 Garden St., Cambridge, MA 02138}

\author{Yuri Aikawa}
\affil{Center for Computational Sciences, University of Tsukuba, 1-1-1 Tennodai, Tsukuba 305-8577, Japan}

\author{Sean M. Andrews}
\affil{Harvard-Smithsonian Center for Astrophysics, 60 Garden St., Cambridge, MA 02138}

\author{Chunhua Qi}
\affil{Harvard-Smithsonian Center for Astrophysics, 60 Garden St., Cambridge, MA 02138}

\author{Ewine F. van Dishoeck}
\affil{Leiden Observatory, Leiden University, PO Box 9513, 2300 RA, Leiden, The Netherlands}
\affil{Max-Planck-Institut f\"ur Extraterrestrische Physik (MPE), Giessenbachstrasse 1, 85748, Garching, Germany}

\author{David J. Wilner}
\affil{Harvard-Smithsonian Center for Astrophysics, 60 Garden St., Cambridge, MA 02138}

\begin{abstract}
In a protoplanetary disk, a combination of thermal and non-thermal desorption 
processes regulate where volatiles are liberated from icy grain mantles into 
the gas phase.  Non-thermal desorption should result in volatile-enriched gas in disk-regions where complete freeze-out is otherwise expected.
We present ALMA observations of the disk around the young star IM Lup in 
1.4~mm continuum, C$^{18}$O 2--1, H$^{13}$CO$^+$ 3--2 and DCO$^+$ 3--2 
emission at $\sim$0\farcs5 resolution. The images of these dust and 
gas tracers are clearly resolved.  The DCO$^+$ line exhibits a striking pair 
of concentric rings of emission that peak at radii of $\sim$0\farcs6 and 
2\arcsec\ ($\sim$90 and 300\,AU, respectively). Based on disk chemistry model comparison, the inner 
DCO$^+$ ring is associated with the balance of CO freeze-out and thermal desorption due to a radial decrease in disk temperature. The outer DCO$^+$ ring is explained by 
non-thermal desorption of CO ice in the low-column-density outer disk, repopulating the disk midplane with cold CO gas. The CO gas then reacts with abundant H$_2$D$^+$ to form the observed DCO$^+$ outer ring.  These observations demonstrate 
that spatially resolved DCO$^+$ emission can be used to trace otherwise hidden cold gas reservoirs in the outmost disk regions, opening a new window onto their chemistry and kinematics.
\end{abstract}


\section{Introduction}

Planets form through the assembly of refractory dust, volatile ice, and 
gas in disks around young stars.  The spatial distribution of volatile 
abundances helps determine the compositions of nascent planets that form at 
different disk radii.  The separation of these volatiles into their gas and 
ice phases is set by the balance of adsorption (``freeze-out" of gas 
onto solids), and desorption back to the gas phase. Desorption results 
from both thermal and non-thermal processes that can be active in different 
regions of protoplanetary disks. 

The thermal desorption (sublimation) rate for a volatile is set by its energy
barrier for ice desorption in concert with the grain temperature. Common 
volatiles like H$_2$O, CO$_2$, and CO are characterized by very different 
desorption barriers. For a typical disk with midplane temperatures that 
increase closer to the host star, this range of desorption energy barriers 
will introduce a series of distinct locations -- ``snow lines" -- where 
specific volatiles are liberated from ices into the gas phase.  The highly 
volatile species CO and N$_2$ have the lowest thermal desorption barriers, 
and so have snow lines at large disk radii where the disk is cold, 
$\sim$30--100\,AU \citep{Oberg11e}.  
Irradiation heating maintains a thermal inversion in the disk atmosphere, 
where temperatures increase with height ($z$) above the midplane 
\citep[e.g.][]{Calvet91}.  These higher surface temperatures mean that 
snow lines are really two dimensional snow surfaces whose boundaries extend 
to larger $r$ for larger $z$.

In addition to thermal desorption, there are analogous non-thermal processes
driven by interactions between ice and high-energy photons and particles, or the release 
of chemical energy \citep{Garrod07}.  Ice photodesorption by ultraviolet (UV)
radiation has garnered special interest because of the strong UV fields 
produced by young stars \citep[e.g.,][]{Bergin03} and the high UV desorption 
yields measured in laboratory experiments 
\citep{Westley95,Oberg07b,Fayolle11,Chen14}.  
In the inner disk, UV photodesorption can result in lower scale heights for 
snow surfaces compared to the ones expected from thermal desorption 
\citep{Woitke09,Hogerheijde11,Oka12}.  At the low densities of the outer disk, 
UV radiation may penetrate to the cold midplane to desorb CO ice well outside 
the expected location of the thermal CO snowline \citep{Willacy00}. 

DCO$^+$ is expected to form efficiently in cold regions where CO gas is still present in the gas-phase, whether thermally or non-thermally desorbed, through deuteron transfer from the low-temperature (T$<$50~K) product H$_2$D$^+$ to CO.  Therefore, the 
expectation is that DCO$^+$ emission peaks close to the CO snow surface, i.e. 
at the coldest disk locations where CO is still abundant in the gas-phase if only thermal desorption is considered. Several studies have found an increasing DCO$^+$ abundance with radius 
ascribed to this process \citep{Qi08,Teague15}, and \citet{Mathews13} resolved 
a ring of DCO$^+$ in the disk around the Herbig Ae star HD 163296, which 
seems closely related to the CO snow line.  Their proposed connection between the DCO$^+$ emission morphology and the CO 
snowline is challenged by new disk chemistry models, where significant DCO$^+$ 
can be produced at temperatures above 30~K through pathways that do not involve
H$_2$D$^+$ \citep{Favre15}.  However, this complication does not eliminate 
the fact that DCO$^+$ {\it requires} gas-phase CO to form, nor does it remove 
the expectation that DCO$^+$ is formed in excess if H$_2$D$^+$ is abundant 
(i.e., if the gas is cold).  

In this study, we present ALMA Cycle 2 observations of the IM Lup disk, 
located at a distance of 155 pc \citep{Lombardi08}, in 1.4\,mm continuum,
DCO$^+$ 3--2, C$^{18}$O 2--1 and H$^{13}$CO$^+$ 3--2 line emission (\S2).  
\S3  presents the observed emission morphology of DCO$^+$ and the other species. In \S4 we use a
disk chemistry model, applied to a generic T Tauri 
disk structure, to interpret the observations. We highlight the role of non-thermally desorbed CO
to explain the presence of DCO$^+$ emission in the outer 
regions of the disk in the discussion in \S5 and summarize our findings in \S6.

\section{Observations and Data Reduction\label{alma}}

\begin{deluxetable*}{r c c c c  c }
\tablecolumns{7}
\tablewidth{0pc}
\tablecaption{Molecular line data\label{lines}}
\tablehead{
\colhead{Molecular Line}   & \colhead{Rest freq.}    & \colhead{Log$\rm_{10}(A_{ij})$	} &
\colhead{E$_{\rm u}$}    & \colhead{beam (PA)}	 & \colhead{Integrated flux}\\
\colhead{}   & \colhead{GHz}    & \colhead{} &
\colhead{K}    &\colhead{$\arcsec\times\arcsec$ (\degree)} &\colhead{mJy km/s}}
\startdata
DCO$^+$	J=3--2	&216.1126	&-2.62 	&20.7	&$0\farcs65\times0\farcs48$ (-71\degree) &$450\pm45$\\
C$^{18}$O J=2--1	&219.5604	&-6.22 	&15.8	&$0\farcs68\times0\farcs47$ (-60\degree)&$1460\pm150$\\
H$^{13}$CO$^+$ J=3--2	&260.2553	&-2.87 	&25.0	&$0\farcs68\times0\farcs59$ (89\degree)&$484\pm48$
\enddata
\end{deluxetable*}

Observations of the IM Lup disk were acquired with the Atacama Large 
Millimeter/Submillimeter Array (ALMA; project code ADS/JAO.ALMA\#2013.1.00226.S) on 2014 July 6
with 31 antennas (targeting DCO$^+$ and C$^{18}$O) and on 2014 July 17 with 32 
antennas (targeting H$^{13}$CO$^+$).  The antenna separations spanned baselines 
of 15--650\,m.  The total on source integration time was $\sim$21 minutes in 
each execution.  The nearby quasars J1534$-$3526 and J1427$-$4206 were used 
for gain and bandpass calibration, respectively.  Titan was observed to 
calibrate the absolute amplitude scale.

On the July 6 execution, the correlator was configured to process 13 distinct 
spectral windows (SPWs).  The spectral resolution $\delta\nu$ and bandwidth for 12 of the SPW were 
61 kHz and 59 MHz, respectively; the remaining SPW 
had a coarser resolution with a bandwidth of 469 MHz.  The DCO$^+$ 3--2 line 
was located in SPW1, and C$^{18}$O 2--1 in SPW6 (Table \ref{lines}). 
For the July 7 execution, the correlator was configured to process 14 SPWs in a 
different part of the Band 6 spectrum.  All had the same spectral resolution, 
61 kHz; twelve SPWs had a bandwidth of 59 MHz, and the remaining two had a 
bandwidth of 117 MHz.  The H$^{13}$CO$^+$ 3--2 line (Table \ref{lines}) was located in SPW1. 

The visibility data were calibrated by ALMA staff.  The calibration plots were 
inspected, deemed adequate, and no changes to the delivered calibration were 
made. 
Each individual SPW was further phase and amplitude self-calibrated in CASA 
4.2.2.  In each SPW, the continuum was subtracted using line-free channels.  
The fully calibrated visibilities were Fourier inverted and CLEANed using the 
Briggs weighting scheme: the robust parameter was set 0.5 for the continuum and 
1.0 for the spectral lines.  A CLEAN mask was manually generated based on the 
bright C$^{18}$O 2--1 line, and then applied to the H$^{13}$CO$^+$ and DCO$^+$ 
lines.  The CLEANed maps were restored with a synthesized beam with FWHM 
dimensions of $\sim0\farcs7 \times 0\farcs5$.  The 1.4\,mm continuum has an RMS 
noise level of $\sim$0.24\,mJy beam$^{-1}$ (dynamic range limited).  The integrated flux density is 
181$\pm$18 mJy (which includes a 10\%\ flux calibration uncertainty), consistent 
with previous SMA measurements \citep{Oberg11a}.  
The resulting RMS noise level in 0.2 km s$^{-1}$ binned channels is 4-6 mJy.

\section{Observational Results\label{obs}}

\begin{figure*}[htp]
\epsscale{1.0}
\plotone{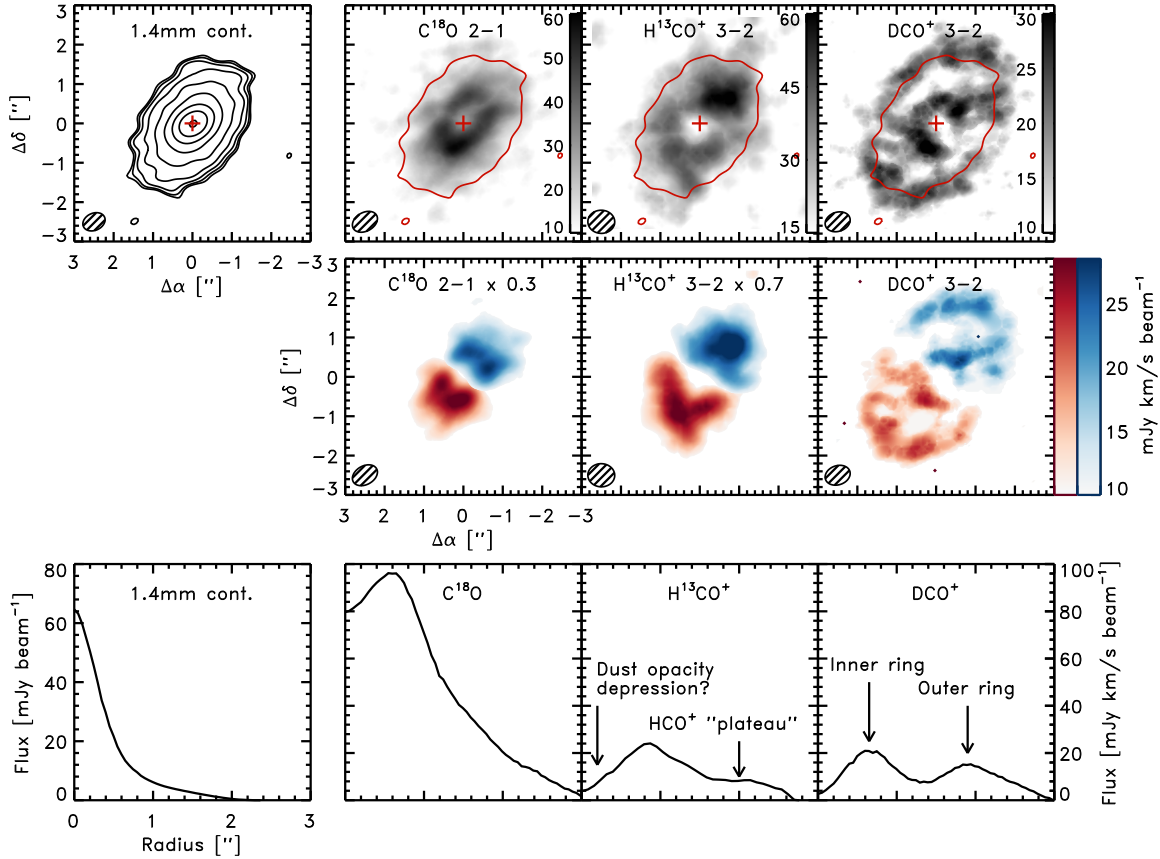}
\caption{Continuum and integrated emission maps of C$^{18}$O 2--1, H$^{13}$CO$^+$ 3--2 and DCO$^+$ 3--2 toward IM Lup (top row), integrated emission in two velocity bins (middle row) and deprojected azimuthally averaged profiles (bottom row). {\it top row:} The contours in the continuum map are 2$\sigma$+[2,4,8,16,...]$\sigma$  with the continuum rms $\sigma=0.24$~mJy. The 2$\sigma$ continuum contour is also shown on top of the integrated line emission maps. The integrated line emission in units of mJy km/s beam$^{-1}$ is shown in gray scale (see individual color bars in each panel). The continuum peak is marked by a red cross and the synthesized beam is displayed in the bottom left of each panel. {\it middle row:} The integrated emission in two velocity bins around the source velocity has been scaled by 0.3 for C$^{18}$O and 0.7 for HCO$^+$ to enable all three line maps to be shown on the same scale (color bar to the right of the DCO$^+$ panel). {\it bottom row:}  The azimuthally averaged molecular emission profiles have been labeled to highlight the lack of molecular emission at the continuum maximum, and the clear DCO$^+$ double-ring structure. \label{fig1}}
\end{figure*}

Figure \ref{fig1} shows the continuum and integrated line emission maps of 
C$^{18}$O 2--1, H$^{13}$CO$^+$ 3--2, and DCO$^+$ 3--2 from the IM Lup disk.  
The continuum is centrally peaked and spatially resolved, extending out to a 
radius of $\sim$2\arcsec\ (300~AU), comparable to previous lower resolution 
observations \citep{Pinte08,Panic09,Oberg11a}.  The peak flux density at the 
disk center is 66$\pm$7 mJy beam$^{-1}$ at 1.4 mm (C$^{18}$O and DCO$^+$ spectral 
setting) and 78$\pm$8 mJy beam$^{-1}$ at 1.1 mm (H$^{13}$CO$^+$ spectral setting), 
corresponding to a brightness temperature of $\sim$6~K.

Integrated line emission maps (Fig. \ref{fig1}, top panel) were constructed by summing up emission from 
individually CLEANed 0.2 km s$^{-1}$-wide channels, employing a 2$\sigma$ 
threshold to maximize the SNR. They are shown together with the continuum 2$\sigma$ contour. The middle panel of Fig \ref{fig1} presents integrated line emission maps of the same lines split up into two velocity bins around the source velocity to visualize the disk rotation. The integrated emission in these panels have been scaled to a constant emission maximum to show the increasing importance of extended emission for DCO$^+$ compared to H$^{13}$CO$^+$, and for H$^{13}$CO$^+$ compared to C$^{18}$O .
To visualize the radial structure of these 
tracers more clearly and to mitigate any bias introduced by the applied 
clipping threshold, we also produced moment-0 maps without any clipping, deprojected and 
azimuthally averaged them into radial brightness profiles 
(Fig. \ref{fig1}, bottom panels).  We assumed an inclination of 49\degr\ and a 
position angle of -35\degr\ for these calculations \citep{Pinte08,Panic09}.  

The integrated emission maps and 
average radial profiles reveal two significant morphological features.  First, 
all three molecular species show a clear central depression.  The C$^{18}$O and
DCO$^+$ emission each peak at a radius of 0\farcs6, or $\sim$90\,AU, and the 
H$^{13}$CO$^+$ line peaks at 0\farcs8 ($\sim$125\,AU).  Second, DCO$^+$ exhibits
a large second ring at 2\arcsec, near the outer edge of the continuum 
emission.  No such ring is observed in C$^{18}$O, H$^{13}$CO$^+$ or the 
continuum.  However, the H$^{13}$CO$^+$ emission ``plateaus'' around the same 
radius and C$^{18}$O exhibits a subtle slope change.

\section{Disk Chemistry Modeling  \label{model}}

\subsection{Model Description}

To interpret the observed C$^{18}$O, H$^{13}$CO$^+$, and DCO$^+$ radial 
structures, we explored the relevant predictions of the disk chemistry model 
presented by \citet{Furuya14,Aikawa15}, updated with deuterium chemistry that considers 
nuclear spin (ortho/para) chemistry of light species. This model adopts
a generic steady, axisymmetric Keplerian disk around a T~Tauri star with 
stellar mass, radius, and effective temperature of $M_{\ast}=0.5$ $M_{\odot}$, 
$R_{\ast}=2$ $R_{\odot}$, and $T_{\ast}=4000$ K, respectively, and disk mass
$1.7\times10^{-2}$ M$_\odot$ \citep{Nomura07}.  The stellar UV and X-ray 
luminosities are set to $10^{31}$ erg s$^{-1}$ and $10^{30}$ erg s$^{-1}$, 
respectively, based on the observed spectrum of the young star TW Hya 
\citep{Herczeg02,Kastner02}. The disk model assumes that the grains have grown up to 1~mm throughout the disk, resulting in a factor of 10 decrease in grain surface area compared to ISM dust. The gas temperature, dust temperature and density distributions of the disk are calculated self-consistently, 
considering various heating and cooling mechanisms \citep{Nomura07}. 

Compared to IM Lup, 
this model employs a fainter star, which results in a colder disk, and also 
a lower disk mass. The differences in temperature and density structure
clearly precludes any quantitative comparison between the model and data. 
However, even though the model has not been tuned to the parameters of 
IM~Lup, qualitative comparisons still provide valuable insight into the 
origins of the observed molecular emission structure, as long as the model results are robust to the model parameter choices. To evaluate the latter we performed a small parameter study where we varied the CO binding energy, the disk density and temperature structure, and the cosmic ionization rate. As described below the main results are not sensitive to these parameters. We therefore conclude that the predictions from the generic disk model are useful for interpreting the observed emission pattern of the specific disk around IM Lup.

The chemical composition of the model is calculated by integrating a 
time-dependent system of gas and grain surface rate equations for 300\,kyr, 
assuming the physical structure is static \citep{Hasegawa92}. 
The gas-grain reaction network is based on the work by \citet{Garrod06}, 
supplemented with calculations that account for high-temperature gas phase 
reactions \citep{Harada10}, X-ray mediated chemistry \citep[][and 
references therein]{Furuya13}, deuterium chemistry \citep[][and references 
therein]{Aikawa12}, and nuclear spin state chemistry of H$_2$, H$_3^+$ and 
their isotopologues \citep{Hugo09,Coutens14}.  
Species containing chlorine, phosphorus, or more than four carbon atoms were 
excluded for computational expediency.  Elemental abundances are 
taken from \citet{Aikawa01}.  The initial abundances are obtained by 
calculating the molecular evolution of a star-forming core, following 
\citet{Aikawa12}. 

Importantly for this study, the binding energy of CO in the generic disk model is set to 1150 K, 
its measured value for H$_2$O ice surfaces \citep{Collings04}, which results 
in a CO midplane freeze-out temperature of $\sim$26~K. If CO is instead freezing out on a pure CO ice, the binding energy will be lower. We explored the effects of such a scenario by running a model with the CO binding energy set to 1000~K. Non-thermal desorption 
is included through stochastic heating by cosmic-rays \citep{Hasegawa93}, 
photodesorption \citep{Oberg09a,Oberg09b,Fayolle11,Chen14}, 
and reactive desorption \citep{Garrod07}. The CO photodesorption yield is set 
to $7\times10^{-3}$ \citep{Fayolle13}. UV photodesorption and cosmic ray
heating are the most important non-thermal desorption mechanisms. Their relative contributions 
depend on a combination of adopted yields, cosmic ray ionization 
rate, dust properties, and UV field, which varies across the disk.

The main target of this study, DCO$^+$, forms through three different pathways in the model 
\begin{enumerate}
\item H$_2$D$^+$+CO$\rightarrow$ H$_2$+DCO$^+$, 
\item HCO$^+$+D$\rightarrow$DCO$^+$+H 
\item CH$_2$D$^+$+O$\rightarrow$DCO$^+$+H$_2$ and other pathways related to 
CH$_2$D$^+$. 
\end{enumerate}
The D and CH$_2$D$^+$ pathways regulate DCO$^+$ formation at elevated 
temperatures, i.e. in the inner (R$<$10~AU) disk and in the disk atmosphere 
at all radii. The D pathway is generally the more important of the two in the present model.
\citet{Favre15}, on the other hand, argued that the CH$_2$D$^+$ path is more important, probably due to a different 
disk ionization structure compared to our model. 
Exterior to 10~AU, warm DCO$^+$ is a minor contribution to the total DCO$^+$ 
column density. Instead, the H$_2$D$^+$ pathway dominates the 
DCO$^+$ production. H$_2$D$^+$ forms efficiently at low temperatures due to 
the small ($<$230~K) zero-point energy difference between H$_3^+$ and 
H$_2$D$^+$ (the exact energy difference depends on the spin states of H$_2$ 
and H$_2$D$^+$).

\subsection{Model Results and Comparison to Observations\label{model_results}}

\begin{figure*}[htp]
\epsscale{1.0}
\plotone{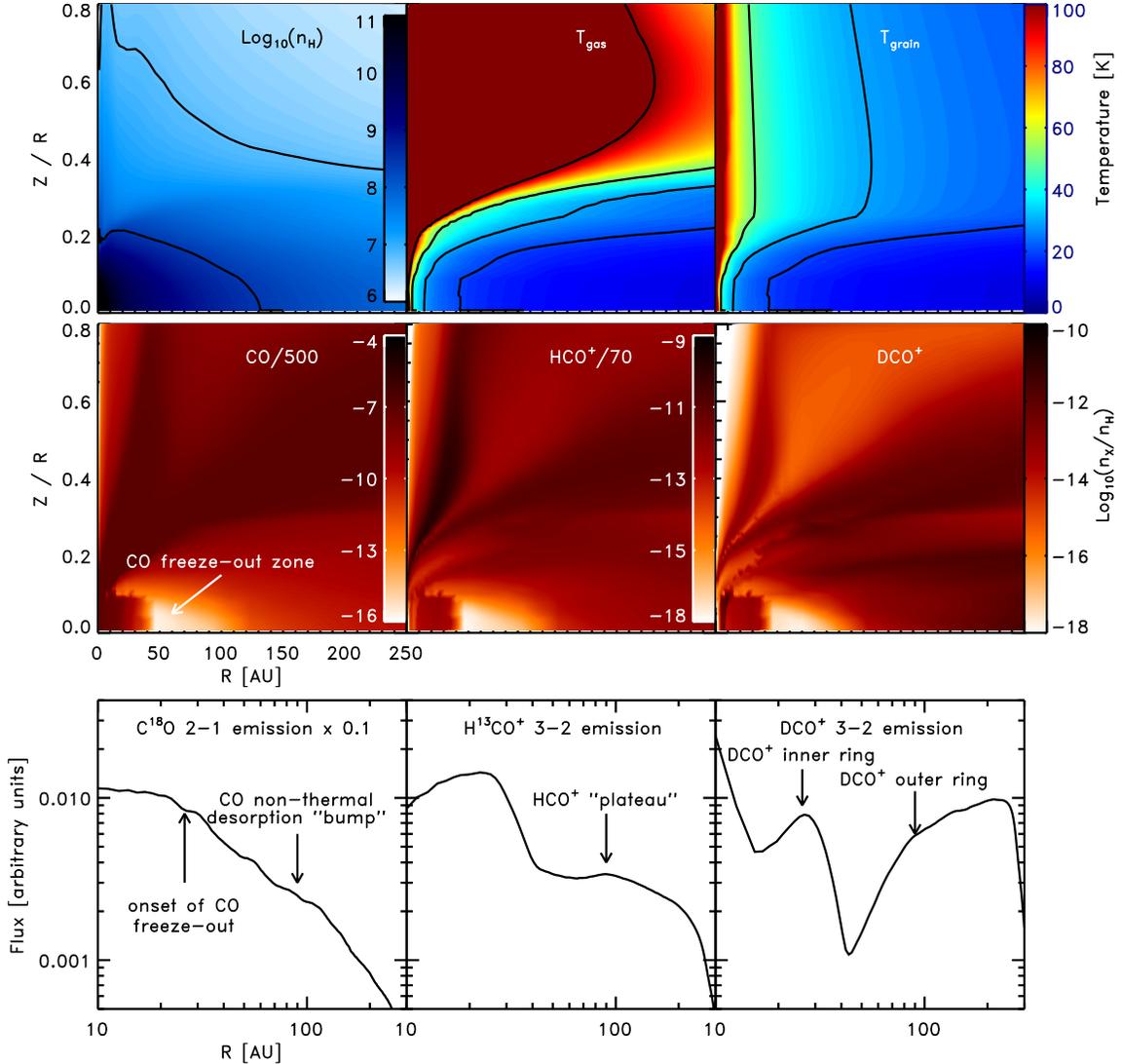}
\vspace{-5mm}
\caption{Density, gas and dust grain temperatures in our generic disk model (top row) and the CO/500 ($\sim$C$^{18}$O), HCO$^+$/70 ($\sim$H$^{13}$CO$^+$) and DCO$^+$ abundances with respect to H nuclei after 300kyr (middle row). The density contours mark $n_{\rm H}$=10$^7$ and 10$^9$ cm$^{-3}$ (upper left panel). The temperature contours mark 20, 30, 50 and 100~K (upper middle and right panels). The bottom row shows the calculated radial emission profiles of the observed lines C$^{18}$O 2--1, H$^{13}$CO$^+$ 3--2 and DCO$^+$ 3--2. \label{fig2}}
\end{figure*}

Figure \ref{fig2} shows the model disk temperature and density structure.
One particular difference worth highlighting between this generic model 
and IM~Lup is in the disk midplane temperature profile. In the model, 
20~K is reached at a radius of $\sim$40~AU, while a previous analysis of the 
IM~Lup disk suggests that 20~K is reached at a larger radius  
($\sim$40--100~AU), dependent in detail on the model assumptions 
\citep{Pinte08}\footnote{We have corrected for the assumed distance of 
190\,pc}. As a result, CO freeze-out in this model probably starts at smaller 
radii compared to the IM Lup disk.

These model calculations do not include C and O isotope-specific chemistry. 
To make comparisons with the observations of isotopes, we have scaled the 
calculated CO and HCO$^+$ densities by factors of 500 and 70, respectively, 
adopting the standard ISM $^{16}$O/$^{18}$O and $^{12}$C/$^{13}$C, to 
estimate the C$^{18}$O and H$^{13}$CO$^+$ densities. We expect that this is a reasonable approximation in the disk midplane (in the disk atmosphere, isotopologue specific photodissociation rates will change the relative CO isotopologue abundances \citep{Miotello14}). 

Figure \ref{fig2} (middle row) shows the radial and vertical gas-phase abundance 
profiles for the species that are the focus of this study, i.e. CO, HCO$^+$ and
DCO$^+$, with the first two abundances scaled to correspond to the observed isotopologues. All three molecules exhibit  
an abundance depression in the disk midplane exterior to 25~AU due to freeze-out 
of CO onto grain surfaces (i.e., the midplane CO snow line). Between 25 and 50~AU, the CO abundance decreases by seven orders of magnitude, the HCO$^+$ and DCO$^+$ abundance by five orders of magnitude. 
 
 Exterior to 90~AU, the CO, HCO$^+$ 
and DCO$^+$ gas-phase midplane abundances begin to increase. With 
decreasing density, the freeze-out rate decreases, which changes the 
CO desorption/freeze-out balance and therefore increases the gas/solid ratio. 
The accompanying decrease in column density results in increased UV penetration
and therefore CO ice UV photodesorption, further increasing the CO gas/solid 
ratio. The result is a substantial amount of CO gas in the cold, outer disk 
midplane, corresponding to a C$^{18}$O abundance of $\sim10^{-11}$ with respect to n$_{\rm H}$. This is six orders of magnitude above the abundance minimum found at 50~AU, but still small compared to the total C$^{18}$O abundance of $\sim10^{-7}$, i.e. most CO (isotopologues) are present in the form of ice in the outer disk midplane. By contrast, a C$^{18}$O abundance of $\sim10^{-7}$ is reached in the warm molecular disk layer, $Z/R\sim 0.3-0.5$, and this is the main reservoir of C$^{18}$O gas in the outer disk. 

HCO$^+$ and DCO$^+$ are also present in the warm molecular disk layer, but this layer is no longer the dominant reservoir. In the case of DCO$^+$, the midplane and molecular layer abundances are comparable. The relatively high DCO$^+$ abundance in the midplane, at $Z/R<0.2$, is due to the increased efficiency of DCO$^+$ production at low temperatures. Because of an increasing gas density toward the disk midplane, the vertically integrated DCO$^+$ column density is dominated by the midplane DCO$^+$ in the outer disk.  HCO$^+$ presents intermediate trends between CO and DCO$^+$. 

We calculated disk emission spectral data cubes using the 
3D Monte Carlo radiative transfer code LIME \citep{Brinch10}  from the 
model disk density, temperature and molecular abundance structures, for the observed C$^{18}$O, H$^{13}$CO$^+$ and DCO$^+$ lines. 
Figure \ref{fig2} (bottom row) shows the radial emission profiles (arbitrary 
units) of the C$^{18}$O, H$^{13}$CO$^+$ and DCO$^+$ obtained from the model
spectral-image data cubes, integrated in velocity space and azimuthally 
averaged like the data. The C$^{18}$O emission profile is centrally peaked. It shows 
a radial gradient small change at $\sim$25\,AU at the onset of CO freeze-out. The effect of CO gas in the midplane
beyond 90~AU on the CO emission profile is subtle, and in practice 
unobservable, since the overall column density and thus emission at these radii
is dominated by the abundant gas-phase CO in the warmer atmosphere layers.   

The HCO$^+$ and DCO$^+$ emission profiles decrease rapidly at radii of 
25--40\,AU, due to the freeze-out of gas-phase CO (their formation reservoir). 
The HCO$^+$ emission then presents a ``plateau'' out to 100~AU, attributed to 
Cosmic Ray and UV photodesorption of CO back into the gas phase. The effect of
the midplane CO non-thermal desorption on the DCO$^+$ emission is, as expected, more dramatic. The return of CO to the gas at large radii, where low temperatures drives ann efficient deuterium fractionation chemistry, results in a second 
maximum (outer ring) peaking at 200~AU. DCO$^+$ also presents a small peak 
toward the central star, caused by warm deuterium fractionation chemistry, 
and an inner ring close to the CO midplane snow line.  Unlike for CO emission, the 
presence of cold CO in the outer midplane has a considerable effect on HCO$^+$ and DCO$^+$ emission, which should be observable.

Based on a small grid of models the shapes of the predicted CO, HCO$^+$ and DCO$^+$ radial profiles are not sensitive to specific model assumptions. In particular the two concentric DCO$^+$ rings are reproduced when changing the CO binding energy from 1150 to 1000~K, i.e. from a water to a CO dominated ice environment, and when the CR ionization rate is reduced to 10$^{-20}$ s$^{-1}$. In the latter case UV photodesorption still maintains cold CO gas in the outer disk, albeit at a lower level compared to when a high cosmic ray ionization rate is assumed. That is, the outer ring is still present when the cosmic ray ionization is attenuated, but it may be less pronounced. The molecular emission trends in Fig \ref{fig2} thus appears to be robust as long as some non-thermal desorption process is active in the disk midplane. However, the absolute DCO$^+$ column density and predicted emission profile in the outer disk do depend on model parameters, and whether the predicted DCO$^+$ double-ring will be observable may thus vary from disk to disk.

Qualitatively, the observed and predicted CO, HCO$^+$ and DCO$^+$ radial 
profiles agree well. In particular, the model reproduces several key features
of the IM~Lup observations: the centrally peaked C$^{18}$O emission profile, the 
H$^{13}$CO$^+$ shoulder in the outer disk, and the DCO$^+$ inner ring and outer ring.
However, the model shows no signs of a central depression for C$^{18}$O, and
only a very small depression for H$^{13}$CO$^+$.  For DCO$^+$, the model 
double-ring is accompanied by an emission peak at the disk center, which is 
also not observed.

\section{Discussion\label{disc}}

\subsection{The DCO$^+$ Outer Disk Ring}

The most surprising result from the ALMA observations is the clear detection of 
a concentric pair of emission rings in the DCO$^+$ 3--2 spectral line, peaking 
at radii of $\sim$90 and 300\,AU.  Based on the midplane temperature profile 
for the IM Lup disk that was inferred from previous measurements 
\citep{Pinte08,Panic09}, the inner ring may coincide with the onset of CO 
freeze-out. The relationship between the CO midplane snowline and the 
inner DCO$^+$ ring location is, however, complicated by multiple DCO$^+$ formation pathways in the inner disk. 

The outer DCO$^+$ ring requires the return of some CO into the gas-phase at large disk radii. This can be accomplished by either non-thermal ice desorption or a radial thermal inversion in the low density outer region. Of the two scenarios we favor the former, since it naturally explains the large DCO$^+$ enhancement and the much smaller H$^{13}$CO$^+$ enhancement at large disk radii. This could be observationally checked by observing additional DCO$^+$ and H$^{13}$CO$^+$ lines that constrain the gas temperature.
 
The location of the outer DCO$^+$ ring coincides with the disappearance of 
mm-grain emission (Fig. \ref{fig1}). It may also coincide with a steepening of the 
gas-density profile \citep{Panic09}. The radial profile of smaller grains, 
which carry most of the opacity at UV wavelengths, is not known independently, 
but it seems reasonable to assume that it too is steepening at the same radius.
The  DCO$^+$ outer ring then likely coincides with a disk location where the 
UV radiation penetration depth is increasing rapidly with radius. The 
connection between the DCO$^+$ outer ring and UV radiation is supported by 
chemical modeling. An outer DCO$^+$ ring is produced when CO is 
non-thermally desorbed from icy grains in the cold, low-density outer disk. 
In the model UV photodesorption and CR heating are both important, but 
the coincidence between the ring and the loss of dust opacity suggests that 
in the case of IM Lup, UV photodesorption is the most important desorption 
mechanism in the outer disk. Once in the gas phase, CO can be chemically 
converted into DCO$^+$ following deuteron transfers from abundant reservoirs 
of H$_2$D$^+$ and other deuterated H$_3^+$ isotopologues (which are enhanced 
at such low temperatures). This effect is expected to be generally present in disks and especially pronounced in disks with significant grain growth, which reduces UV opacity. In our model we assume a uniform grain growth throughout the disk, parameterized as a reduction in total grain surface area by a factor of 10. Further grain growth is likely present in older disks, and it may therefore be illustrative to compare the DCO$^+$ distribution in disks of different ages and with different grain properties as estimated from continuum observations \citep{Perez12, Andrews14}. 

Based on the discussion above and the seemingly robust model results in \S\ref{model_results}, it is tempting to conclude that concentric DCO$^+$ rings should be commonly observed in protoplanetary disks. It is important to realize however, that the expected intensity of the second DCO$^+$ ring {\it is} model dependent. \citet{Teague15}, for example, finds a DCO$^+$ enhancement in the outer disk, but it is small compared to the findings in our model. The same is true in the model by \citet{Willacy07}. 
In general, the DCO$^+$ abundance in the outer disk will depend on the intensity and penetration depths of radiation fields (UV, cosmic rays and X-rays), and the DCO$^+$ ring column density will depend on the total amount of cold molecular gas exposed to this radiation. The intensity of the DCO$^+$ ring will also depend on the excitation conditions that characterize the outer disk midplane, especially its density and temperature. It may very well be that DCO$^+$ rings, while common, are only observable in a small sub-set of disks where disk density, temperature and radiation structures conspire to produce sufficient amounts of DCO$^+$ flux at large disk radii.  Still the detection of one such outer ring in the IM~Lup disk demonstrates that non-thermal desorption in general, and UV photodesorption of CO in particular, is important in setting the chemical structure of protoplanetary disks. 

\subsection{Origins of the Central Depression in Several Tracers}

A lingering puzzle in the observations presented here is the central 
depression in the intensities of all the observed spectral lines (but notably 
not in the continuum). We consider four potential explanations for these 
depressions: 
(1) a reduction in the gas surface densities; 
(2) line-of-sight absorption by foreground cloud material; 
(3) unexpected chemistry; and 
(4) high continuum optical depths that prevent some of the line emission 
from escaping the disk. 

The first possibility seems unlikely, since there is no 
analogous signature in the continuum, and we are unaware of a physical 
mechanism that would systematically enhance the mm dust-to-gas ratio in the inner 
disk.  The second scenario, molecular cloud contamination, should produce deeper 
depressions (absorption) for the more optically thick spectral lines.  
Inspection of CO data toward the same source show that CO
exhibits less visible depressions than C$^{18}$O, ruling out this scenario.

Chemistry can produce central depressions 
in HCO$^+$ and DCO$^+$ following desorption of 
H$_2$O and other molecules with high proton affinity in the inner disk, 
resulting in efficient proton transfer from HCO$^+$ and DCO$^+$ to H$_2$O.  Chemistry may 
thus contribute to the H$^{13}$CO$^+$ and DCO$^+$ holes. The C$^{18}$O central 
depression is not predicted by our or any other published disk chemistry model 
\citep[e.g.][]{Willacy09,Walsh10,Semenov11}, though the conversion efficiency 
of CO into other species across the disk is an active topic of research 
\citep[e.g.][]{Favre13}.  Based on the results presented so far, it seems 
unlikely, however, that chemistry alone can explain the entire observed effect.

The last proposed explanation is that the continuum optical depths in the 
disk center are high enough to block some of the line emission from the midplane.  The deeper depressions observed for more optically thin tracers, which originate closer to the midplane, and for lines emitted at higher frequencies (especially H$^{13}$CO$^+$) supports this scenario. But this scenario has one significant problem: the peak continuum 
intensity at the disk center corresponds to a brightness temperature of only 
6~K, much lower than expected for optically thick emission at these radii.  
The only way of reconciling this explanation with the data is through 
significant beam dilution.  If the optically thick continuum emission is 
concentrated in small scale features (occupying an area $\sim$5$\times$ smaller
than the synthesized beam), then this scenario may work.  That may seem 
unlikely, but the recent observations of the HL Tau disk showing small scale
structure \citep{Brogan15} indicate that it remains a possibility.  This 
hypothesis could be tested directly by very high spatial resolution continuum 
measurements, or less directly by observations of CO and HCO$^+$ isotopologue 
lines at 3~mm, where the dust opacity should be reduced.

\section{Summary and Conclusions\label{sum}}

ALMA observations of the massive disk around IM Lup have revealed a spectacular
DCO$^+$ double-ringed structure. The inner ring appears connected with the thermal CO snowline, but dust opacity and multiple DCO$^+$ formation pathways  complicate the interpretation. The second, outer DCO$^+$ emission ring at 300~AU
is readily reproduced by non-thermal CO ice desorption in the low-density, outer disk. The ring coincides with a rapid decrease in dust opacity, which is likely accompanied by increased UV penetration toward the disk midplane. We hypothesize that the DCO$^+$ outer ring is caused by efficient UV photodesorption of CO ice at this radius.

Without the presence of non-thermal desorption this outer disk region would be completely devoid of CO gas and thus of CO-gas-mediated chemistry. Even at its most efficient, non-thermal desorption in disk mid planes is not expected to maintain a large fraction of CO in the gas-phase, $<0.1$\% of the total CO reservoir. The presented observations and models show, however, that the release of such trace amounts of CO into the gas-phase can still produce an observable signature through its effects on DCO$^+$ production. This DCO$^+$ can be used as a probe of the kinematics and ionization in the outer disk midplane, opening up a new window on the structures of protoplanetary disks.

\acknowledgments

We are grateful to Ugo Hincelin and Eric Herbst for providing rates and branching ratios for ortho/para chemistry.
This paper makes use of the following ALMA data:
ADS/JAO.ALMA\#2013.1.00226.S. ALMA is a partnership of ESO (representing
its member states), NSF (USA) and NINS (Japan), together with NRC
(Canada) and NSC and ASIAA (Taiwan), in cooperation with the Republic of
Chile. The Joint ALMA Observatory is operated by ESO, AUI/NRAO and NAOJ. The National Radio Astronomy Observatory is a facility of the National Science Foundation operated under cooperative agreement by Associated Universities, Inc. KI\"O also acknowledges funding from the Alfred P. Sloan Foundation, and the Packard Foundation.  K.F. is supported by a Research Fellowship from the Japan Scoiety for the Promotion of Science (JSPS). Astrochemistry in Leiden is by
 A-ERC grant 291141 CHEMPLAN and a KNAW professorship prize.

{\it Facilities:} \facility{ALMA}.

\bibliographystyle{aa}

\end{document}